\documentclass[aps,pra,twocolumn,superscriptaddress,showpacs,amsmath,amssymb,nofootinbib]{revtex4-1}
\usepackage{graphicx}
\usepackage{dcolumn}
\usepackage{bm}

\usepackage{color}

\newcommand{\beq}{\begin{equation}}
\newcommand{\eeq}{\end{equation}}
\newcommand{\beqa}{\begin{eqnarray}}
\newcommand{\eeqa}{\end{eqnarray}}

\newcommand{\ignore}[1]{}
\newcommand{\mostra}[1]{}

\begin{document}

\title{Thermal spin fluctuations in spinor Bose-Einstein condensates}
\date{\today}

\author{M. Mel\'e-Messeguer}
\affiliation{Departament d'Estructura i Constituents de la Mat\`{e}ria,
Universitat de Barcelona, 08028 Barcelona, Spain}

\author{B. Juli\'a-D\'iaz}
\affiliation{ICFO-Institut de Ci\`encies Fot\`oniques, 
Parc Mediterrani de la Tecnologia, 08860 Barcelona, Spain}

\author{A. Polls}
\affiliation{Departament d'Estructura i Constituents de la Mat\`{e}ria,
Universitat de Barcelona, 08028 Barcelona, Spain}

\author{L. Santos}
\affiliation{Institut f\"ur Theoretische Physik, Leibniz 
Universit\"at Hannover, 30167 Hannover, Germany}

\begin{abstract}
We study the thermal activation of spin fluctuations in dynamically-stable 
spinor Bose-Einstein condensates. We analyze the specific cases of a non-dipolar spin-$1$ condensate in $m=0$, where thermal activation results from 
spin-changing collisions, and of a Chromium condensate in the maximally stretched state $m=-3$, where thermal spin fluctuations are due to 
dipole-induced spin-relaxation. In both cases, we show that the low energy associated to the spinor physics may be employed for thermometry purposes 
down to extremely low temperatures, typically impossible to measure in BECs with usual thermometric techniques.
Moreover, the peculiar dependence of the system's entropy with the applied Zeeman energy opens a possible route for adiabatic cooling.
\end{abstract}

\maketitle


\section{Introduction}
\label{sec:Introduction}

Ultra-cold spinor gases, formed by atoms with multiple internal states, have attracted a major interest 
due to the rich physics resulting from the interplay between 
internal and external degrees of freedom~\cite{Kawaguchi2012}. In particular, spinor Bose-Einstein condensates~(BECs)
have been throughly investigated for gases of different spins, including spin-$1$~($F=1$ $^{87}$Rb~\cite{Chang2004} and  
$^{23}$Na~\cite{Stenger1998}), spin-$2$ ($F=2$ $^{87}$Rb~\cite{Schmaljohann2004}), and spin-$3$~($^{52}$Cr~\cite{Fattori2006,Pasquiou2011}).

Especially interesting is the possibility of a coherent population transfer between the different internal 
components. This is accomplished through the so-called 
spin-changing collisions, which, although conserving the total spin projection of the colliding pair of atoms, induce a redistribution 
among the various internal levels~\cite{Kawaguchi2012}. These collisions originate from the differences between intra- and inter-component scattering lengths. 
Since these differences are very small in usual experiments, spin-changing collisions are hence associated to very low energy scales, 
typically much smaller than the overall chemical potential.

The case of Chromium deserves a separate discussion. Chromium is not only an example of a spin-$3$ gas, but presents as well large 
magnetic dipole-dipole interactions~(DDI)~\cite{Lahaye2009}. Contrary to the usual contact-like isotropic interactions which as mentioned above preserve spin projections, 
DDI are anisotropic and hence allow for spin relaxation, a phenomenon which has been explored in Chromium BECs in recent years~\cite{Fattori2006,Pasquiou2011}.
As for the case of spin-changing collisions in non-dipolar spinor BECs, spin relaxation in Chromium is associated as well to very low energy scales.

In this paper, we study spin fluctuations in spinor BECs, and in particular the dependence of these fluctuations on temperature. 
BEC thermometry is typically performed using time-of-flight measurements, either 
from the expansion velocity of the thermal cloud or from a bimodal fitting which allows to establish the ratio between condensate and thermal cloud. 
These techniques fail however for low temperatures for which the thermal population is small compared to the number of particles in the condensate. 
The use of low-energy phase fluctuations for thermometry in a two-well scalar condensate was pioneered in Ref.~\cite{Gati2006}. In this paper, we show 
that the very low energy associated to both spin-changing collisions and spin-relaxation opens as well 
interesting possibilities for thermometry purposes down to extremely low temperatures. 
Moreover, we show that the dependence of the entropy of the gas on the Zeeman energy allows for a possible mechanism for adiabatic cooling.

Our paper is structured as follows. In Sec.~\ref{sec:Spin1} we analyze the case of a stable spin-$1$ BEC prepared in the $m=0$ Zeeman sublevel. 
We study by means of the corresponding Bogoliubov analysis the thermally activated spin fluctuations resulting from spin-changing collisions. 
Sec.~\ref{sec:Spin3} is devoted to the case of a Chromium condensate prepared in the maximally stretched Zeeman state, $m=-3$. This case differs 
significantly from the spin-$1$ case, since  the thermally activated spin fluctuations result from the spin-relaxation induced by the DDI. We show that in both scenarios 
spin fluctuations may be employed for deep temperature thermometry and adiabatic cooling. In Sec.~\ref{sec:Conclusions} we summarize our conclusions.




\section{Spin-$1$  condensate}
 \label{sec:Spin1}

A spin-$1$ BEC is described by the Hamiltonian~\cite{Ho1998}

\beqa
\hat{H} &=& \sum_m \int d\bold{r}\; \hat{\Psi}_m^\dagger 
\Bigg[-\frac {\hbar^2\nabla^2}{ 2M} + V_{\rm trap}  + qm^2 \Bigg] \hat{\Psi}_m  
\nonumber \\ 
&+&  \frac {c_0}{ 2} \sum_{m,m'} \int
d\bold{r} \hat{\Psi}_m^\dagger  \hat{\Psi}_{m'}^\dagger  
\hat{\Psi}_{m'}   \hat{\Psi}_m  
\\
&+& 
\frac{c_2}{ 2} 
\sum_{\begin{subarray}{c}  m_1,m_2\\ m_3,m_4\end{subarray}} 
\int d\bold{r} \hat{\Psi}_{m_1}^\dagger  \hat{\Psi}_{m_2}^\dagger  
\bold{F}_{m_1m_3} \cdot\bold{F}_{m_2m_4} \hat{\Psi}_{m_3} \hat{\Psi}_{m_4} 
\;, \nonumber 
\eeqa
where $\hat{\Psi}_m(\bold{r}) (\hat{\Psi}_m^\dagger(\bold{r}))$ is the 
annihilation (creation) operator of boson in the Zeeman state $m=0, \pm 1$ 
at position $\bold{r}$, $M$ is the  atomic mass, $V_{\rm trap}(\bold{r})$ is the external 
trapping potential, and 
$\bold{F}=(F_x,F_y,F_z)$ is the spin vector operator~($F_{x,y,z}$ are the spin-1 matrices). 
The couplings $c_0=4\pi \hbar^2(a_0 + 2a_2)/3M$ and $c_2=4\pi \hbar^2(a_2 - a_0)/3M$, 
are expressed in terms of the scattering legths $a_{0,2}$ which characterize 
low energy collisions with total spin 0 and 2, respectively. The term 
$q m^2$ denotes the quadratic Zeeman shift in an external magnetic field $B_0$, with 
$q=\mu_B^2B_0^2/8C_{\rm hfs}$, with $C_{\rm hfs}$ the hyperfine coupling strength. 
Note that collisions conserve the total spin projection and hence the linear Zeeman energy 
is a conserved quantity which may be gauged out. 

We assume a stable $m=0$ BEC (we will 
discuss the conditions for stability below). 
Assuming small fluctuations around the condensate solution, the spinor BEC is 
described by the field operator
${\hat{\vec{\Psi}}}=(0,\psi_0,0)+(\delta\hat{\Psi}_{-1},\delta\hat{\Psi}_{0},\delta\hat{\Psi}_{+1})$, 
where the BEC wavefunction $\psi_0$ fulfills the Gross-Pitaevskii equation 
\beqa
\bigg[-{\hbar^2 \nabla^2 \over 2M} + V_{\rm trap}(\bold{r}) 
+ c_0  n_0(\bold{r})\bigg] \psi_0(\bold{r}) = \mu \psi_0(\bold{r}) \; ,
\label{eq:gp}
\eeqa
where $\mu$ is the chemical potential, and $n_0(\bold{r})=|\psi_0(\bold{r})|^2$. Retaining up to second order in the fluctuations, we obtain an effective Hamiltonian for $\delta\hat{\Psi}_{\pm 1}$: 
\beqa
\hat{H}_{1} &=& \sum_{m=\pm1} \int d\bold{r}\; \delta\hat\Psi^\dagger_m H_{\rm eff}
\delta\hat\Psi_m   
\nonumber \\
&+&  c_2 \int d\bold{r}\; n_0   \left [
\delta \hat\Psi_{+1}   \delta \hat\Psi_{-1} +{\rm h.c.}  
\right ] \;,
\label{eq:300}
\eeqa
where $ H_{\rm eff} = - \hbar^2\nabla^2/(2M) + {\cal V}(\bold{r})+q$, with 
${\cal V}(\bold{r}) =  V_{\rm trap}(\bold{r}) + (c_0+c_2)n_0(\bold{r}) - \mu$. 
In homogeneous space, $V_{\rm trap}=0$, $n_0$ is constant and we may move to momentum~($\bold {k}$) space where 
the Hamiltonian becomes:
\beqa
\hat H_1 &=&
\int d\bold{k}\;  \varepsilon_k \sum_{m=\pm 1}
\delta \hat \Psi_m^\dagger (\bold{k}) \delta \hat \Psi_m (\bold{k})
\nonumber \\ &+&
c_2 n_0 \int d \bold{k} 
\left[
\delta \hat \Psi_{+1}(\bold k)\delta \hat \Psi_{-1}(-\bold k) + {\rm h.c.} \right ] \label{eq:301} \; ,
\eeqa
where $\varepsilon_k =  {\hbar^2 k^2\over 2M} + c_2 n_0 +q$. 
Using the symmetric and antisymmetric bosonic operators
$\hat S_{\bold k},\hat A_{\bold k} \equiv {1\over \sqrt{2}} 
\left ( \delta \hat \Psi_{+1}(\bold k) \pm \delta \hat \Psi_{-1}(\bold k) \right )$,
and applying a Bogoliubov transformation
$\hat{S}_{\bold{k}},\hat{A}_{\bold{k}} = r_{\bold{k}} \hat{B}_{\bold{k}}^{(S,A)} + t_{\bold{k}} \hat{B}_{\bold{-k}}^{(S,A)\dagger}$, 
with $r_k^2- t_k^2 =1$, and $\varepsilon_k r_kt_k \pm {c_2 n_0\over 2} \big(r_k^2 + t_k^2\big) = 0$, 
we may re-write:
\beqa
\hat{H}_1 = \int d\bold{k} E_k \Big[
\hat B_{\bold k}^{(S)\dagger}\hat B_{\bold k}^{(S)} 
+ \hat B_{\bold k}^{(A)\dagger}\hat B_{\bold k}^{(A)} \Big] 
\; , \label{eq:ham-b}
\eeqa
where $E_k = \sqrt{\varepsilon_k^2 - c_2^2 n_0^2}$ is the spectrum of spin excitations.
Note that  the $m=0$ BEC is stable as long as $E_k$ is real, 
which demands $q>q_{cr}$, with $q_{cr}=(|c_2|-c_2)n_0$.

In equilibrium at a temperature $T$~($\beta=1/k_BT$ with $k_B$ the Boltzmann constant)
$\langle \hat{B}_{\bold{k}}^{(i)\dagger} \hat{B}_{\bold k}^{(i)} \rangle =
\left ( e^{-\beta E_{\bold{k}}^{(i)}} -1 \right )^{-1} $ with $i=S,A$,  and 
$\langle \hat{B}_{\bold{k}}^{(i)\dagger} \hat{B}_{\bold k}^{(j\neq i)} \rangle = 0$. 
Re-expressing $\delta \hat \Psi_{\pm 1}$ as a function of the quasi-particle excitations, 
we obtain the density in $m=\pm1$ 
\beqa
n_{\pm 1}(T) = {1\over (2\pi)^3} \int d\bold{k} \; n_{\pm 1}(k,T) \label{eq:ndens} \,,
\eeqa
with $n_{\pm 1}(k,T) =\langle \delta \hat \Psi_{\pm 1}^\dag \delta \hat \Psi_{\pm 1}\rangle$
given by
\beqa
n_{\pm 1}(k,T) &=&
{\varepsilon_k \over E_k} {1\over {\rm e}^{\beta E_k}- 1}
+ {\varepsilon_k \over 2 E_k } -{1\over 2} \label{eq:kndens} \,,
\eeqa
the same for both components.

\begin{figure}[t]
\begin{center}
\includegraphics[clip=true,width=0.95\columnwidth]{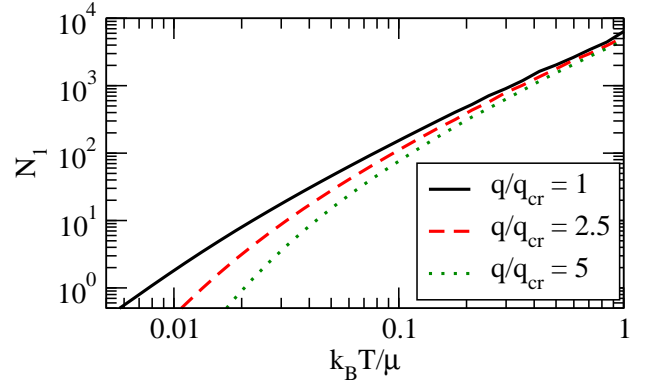}
\caption[]{Ratio $N_{\pm 1}/N_0$ in the trap as a function 
of $q/q_{cr}$ and $k_B T/\mu$ for the case discussed in the text.}  
\label{fig:1}
\end{center}
\end{figure}

We consider at this point a trapped BEC in the Thomas-Fermi~(TF) regime with density profile
$n_0(\bold{r})=\frac{\mu}{c_0} \left ( 1- r_\perp^2/R_\perp^2-z^2/R_z^2\right )$, where $R_{\perp,z}=\sqrt{2\mu/(M\omega_{\perp,z}^2)}$ are the TF radia.
For a sufficiently smooth density profile, we may employ 
the local density approximation~(LDA), associating to each value of the local density $n_0(\bold{r})$ 
the corresponding excitation spectrum for the homogeneous case with that density, $E_k(\bold{r})$. 
We may then evaluate the corresponding local density $n_{\pm 1}(\bold{r})$ from the expressions obtained above.
The total number of atoms in $m=\pm 1$ is obtained by integrating their 
local occupation over the density profile of the trap:
\beqa
\langle N_{\pm 1} \rangle &=& \int d^3r \langle n_{\pm 1}(\bold{r})\rangle \,.
\eeqa
The critical value of the magnetic field, $q_{cr}$ is calculated at the trap center.

We consider in the following the specific case of $F=1$ $^{87}$Rb, 
for which $a_0=101.8 a_B$ and $a_2=100.4 a_B$~(with $a_B$ the Bohr radius).
As a consequence, $c_2=-4.6\times 10^{-3}c_0$ is small and negative, which provides a critical $q_{cr}/\mu= 2 |c_2| / c_0 = 9.25\times 10^{-3}$. For simplicity we consider that the atoms are confined in a spherically symmetric trap, 
with a harmonic frequency $\omega=2\pi \times 50$Hz. For a typical value of $N=10^5$ atoms, the density at the trap center becomes $10^{14}$ cm$^{-3}$.

\begin{figure}[t]
\begin{center}
\includegraphics[width=0.85\columnwidth, clip=true]{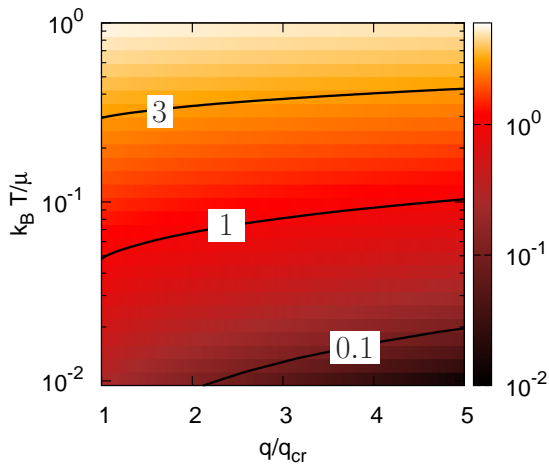}
\caption[]{Entropy as a function of $k_B T/\mu$ 
and $q/q_{cr}$ for the same case of Fig.~\ref{fig:1}. The black curves indicate various iso-entropic curves, the value of $S/k_B$ is indicated.} 
\label{fig:2}
\end{center}
\end{figure}

In Fig.~\ref{fig:1} we plot the total number of atoms 
$\langle N_{\pm 1}\rangle/N_0$ in the trap as a function 
of $k_BT/\mu$ for different values of $q/q_{cr}$. As expected, when approaching the critical $q_{cr}$ 
the spin population is enhanced at low temperatures, due to the very low energy associated to spin excitations. 
At $q=q_{cr}$ the population of $m=\pm 1$~(always for $N=10^5$ atoms in the BEC) is larger than $10$ atoms 
for temperatures larger than $0.02 \mu/k_B$. Although these numbers may be very small to be observed experimentally, 
a simple procedure may significantly enhance the experimental resolution of such small populations. If the system is abruptly 
brought into instability by sweeping into $q<q_{cr}$, spin excitations grow in a process similar to parametric amplification in 
non-linear optics~\cite{Klempt2010}. Within the so-called linear regime, spin fluctuations are hence exponentially amplified with a growth rate $\Gamma$. 
As a result the population in $\pm 1$, is enhanced in time in the form $N_{\pm 1}(t)\simeq N_{\pm 1}(0) e^{\Gamma t}$, where $N_{\pm 1}(0)$ is the 
population in $m=\pm 1$ prior to the destabilization, i.e. that depicted in Fig.~\ref{fig:1}. A subsequent Stern-Gerlach arrangement in time-of-flight~\cite{Klempt2010}, 
allows for a separate imaging of the different Zeeman components. As a result even very small $N_{\pm 1}(0)$ may be 
experimentally resolved, opening the possibility of employing thermally activated spin fluctuations as a thermometer down to temperatures close to $0.01 \mu/k_B$ where 
other thermometry methods typically fail.

Interestingly, the dependence of spin excitations on $q$ may be employed as a possible mechanism for adiabatic cooling. 
For an homogeneous gas, the entropy density of the system is readily calculated from the spectrum of elementary excitations
\beqa
{s\over k_B} &=& \frac{-2}{(2\pi)^3} \int d\bold{k} \bigg[\ln
\left(1-{\rm e}^{-E_k/k_B T} \right)\nonumber \\ 
&&\qquad \quad +  \frac{E_k}{k_B T}\frac{1}{1-e^{E_k/k_BT}} 
\bigg] \,. 
\eeqa
Note that scalar excitations of the $m=0$ condensate, $\frac{\hbar^2k^2}{2M}\left ( \frac{\hbar^2k^2}{2M} + 2c_0n_0 \right )$, 
corresponding to $ \delta\hat{\Psi}_0$ fluctuations, also contribute in principle to the system entropy. However, if $c_2\ll c_0$~(as it is typically the case), 
the entropy contribution of the scalar excitations may be neglected for $k_B T\ll \mu$. In particular, we have checked that for $q$ in the vicinity of $q_c$ 
the entropy contribution of the scalar modes is negligible for $k_BT/\mu<0.25$. 
For a trapped gas, we employ again local density arguments, calculating the local entropy density associated to each position $s(\bold{r})$.  
Fig.~\ref{fig:2} shows curves of equal entropy $S=\int d\bold{r}\, s(\bold{r})$ as a function of temperature and $q/q_{cr}$ for the same case discussed in Fig.~\ref{fig:1}.
Interestingly, the isotropic curves bend to lower temperatures when approaching $q_{cr}$. As a result, an adiabatic variation of the applied magnetic field 
may allow for adiabatic cooling. For example, in the figure, starting with $q=5q_{cr}$ at $T=0.1\mu/k_B$, the system may decrease its temperature down to $0.05 \mu/k_B$
when approaching $q_{cr}$.



\section{Chromium condensate}
\label{sec:Spin3}

We consider in the following the case of a Chromium BEC, which constitutes an example of spin-$3$ gas in which, crucially,  
strong magnetic dipole-dipole interactions~(DDI) induce spin relaxation. As a consequence the low temperature physics is considerably different compared to that of spin-$1$ BECs 
discussed in the previous section. For the spin-$3$ Chromium BEC the Hamiltonian acquires the form $\hat H = \hat H_0 + \hat V_{sr} + \hat V_{dd}$, where 
\beqa
\!\!\!\!\!\hat H_{\rm 0}\!=\! \sum_m\! \int \! d\bold{r}\, \hat \psi_m^\dagger \!
\left[ \frac{-\hbar^2\nabla^2}{2M} \! +\!  V_{\rm trap} \! -\!  \mu\! +\! pm \! +\! qm^2 \! \right ]\!  \hat \psi_m,
\eeqa
where the linear Zeeman energy is included since, contrary to the spin-$1$ case, the global spin projection is not conserved. Note also that 
in principle Chromium presents no quadratic Zeeman effect due to the absence of hyperfine structure. However, an effective quadratic Zeeman effect 
may be induced by optical or microwave dressing~\cite{Gerbier2006,Santos2007}. Contact interactions are 
described by~\cite{Santos2006}:
\beqa
\hat{V}_{sr} &=& {1\over 2} \int d\bold{r} 
:\Big[c_0 \hat{n}^2(\bold{r})+c_1 \hat{F}^2(\bold{r}) \nonumber \\ 
&& \qquad \qquad 
+
c_2 \hat{\cal P}_0(\bold{r}) + c_3 \hat{O}^2(\bold{r}) \Big]: 
\eeqa
where the symbol $::$ denotes normal order; 
$ \hat{n}(\bold{r}) = \sum_m \hat{\psi}_m^\dagger(\bold{r}) \hat{\psi}_m(\bold{r})$ 
is the total density,  
$\hat{F}^i(\bold{r}) = \sum_{m,n} \hat{\psi}_m^\dagger(\bold{r}) S_{m,n}^i \hat{\psi}_n(\bold{r}) $,
$\hat{F}^2(\bold{r}) = \sum_i \big(\hat{F}^i(\bold{r})\big)^2 $, 
$\hat{\cal P}_0(\bold{r}) = {1\over 7} \sum_{m,n} (-1)^{m+n} \hat{\psi}_m^\dagger(\bold{r})
\hat{\psi}_{-m}^\dagger(\bold{r}) \hat{\psi}_n(\bold{r}) \hat{\psi}_{-n}(\bold{r})$, 
$\hat{O}^{ij}(\bold{r}) = \sum_{m,n}\hat{\psi}_m^\dagger(\bold{r}) 
\big(S^i S^j\big)_{m,n} \hat{\psi}_n(\bold{r})$,
$\hat{O}^2(\bold{r}) = \sum_{i,j}\big( \hat{O}_{i,j}(\bold{r})\big)^2 $. 
The interaction constants are 
$ c_0 = (-11g_2 + 81g_4 + 7g_6)/77$, $c_1= (g_6-g_2)/18$,  
$c_2 = g_0 + (-55g_2 + 27g_4 - 5g_6)/33$ and $c_3 = g_2/126 - g_4/77+g_6/198$, 
where as in the previous section, $g_S=4\pi \hbar^2 a_S/M$ for the channel of total spin $S$.
For $^{52}$Cr, $a_6 = 112 a_B$, $c_0 = 0.65 g_6$, $c_1 = 0.059g_6$, $c_2 = g_0 + 0.374g_6$ and  
$c_3 = -0.002g_6 $~(the value of $a_0$ is as yet unknown). Finally, the  DDI acquires the form~\cite{Santos2006}: 
\beqa
\hat{V}_{dd} 
&=& -\sqrt{3\pi\over 10} c_d \iint  {d\bold{r}\,d\bold{r}'\over |\bold{r}-\bold{r}'|^3}\times 
\nonumber \\
&& 
:\bigg[{\cal F}_{z,z}(\bold{r},\bold{r}') Y_{20}(\widehat{\bold{r}-\bold{r'}})
+{\cal F}_{z,-}(\bold{r},\bold{r}') Y_{21}(\widehat{\bold{r}-\bold{r'}})
\nonumber \\ &&
+{\cal F}_{z,+}(\bold{r},\bold{r}') Y_{2-1}(\widehat{\bold{r}-\bold{r'}})
+{\cal F}_{-,-}(\bold{r},\bold{r}') Y_{22}(\widehat{\bold{r}-\bold{r'}})
\nonumber \\ &&
+{\cal F}_{+,+}(\bold{r},\bold{r}') Y_{2-2}(\widehat{\bold{r}-\bold{r'}})
\bigg]:
\eeqa
where $Y_{2m}(\widehat{\bold{r}-\bold{r'}})$ are the spherical harmonics, 
\beqa
{\cal F}_{z,z}(\bold{r},\bold{r}') &=& \sqrt{2\over 3} \Big[3\hat{F}_z(\bold{r})
\hat{F}_z(\bold{r}') - \bold{\hat{F}}(\bold{r}) \cdot \bold{\hat{F}}(\bold{r}')\Big]
\nonumber\\
{\cal F}_{z,\pm}(\bold{r},\bold{r}') &=& \pm \Big[\hat{F}_{\pm}(\bold{r})
\hat{F}_z(\bold{r}') + \hat{F}_z(\bold{r}')\hat{F}_{\pm}(\bold{r})\Big]\nonumber\\
{\cal F}_{\pm,\pm}(\bold{r},\bold{r}') &=&
\hat{F}_{\pm}(\bold{r})\hat{F}_{\pm}(\bold{r'})\,,
\eeqa
and  
$\hat{F}_{\pm}(\bold{r}) = \hat{F}_x(\bold{r}) \pm i\hat{F}_y(\bold{r})$. 
The DDI are characterized by the coupling constant $ c_d = \mu_0 \mu_B^2 g_L^2/ 4\pi$, 
where $\mu_0$ is the vacuum magnetic permeability, $\mu_B$ 
the Bohr magneton and $g_L$ the Land\'e factor. For  $^{52}$Cr,  $g_L=2$ and $c_d=0.004g_6$.  
Note that the DDI do not conserve 
spin and orbital angular momentum separately allowing for spin relaxation processes, in which e.g. 
atoms in $m=-3$ are transferred into $m=-2$~\cite{Santos2006}. 

In the following we consider that the linear and quadratic Zeeman effects are chosen 
in such a way that only the two lowest states of the Zeeman manifold, $m=-3$ and $m=-2$ contribute, 
whereas spin relaxation to other $m$ states is energetically suppressed. In this simplified scenario, we assume 
a condensate of $m=-3$ atoms with small spin fluctuations populating the $m=-2$ component.
This system can be described by the field 
$\hat\psi \simeq  \psi_{-3} + \delta\hat{\psi}_{-3} + \delta\hat{\psi}_{-2}$, where the BEC wavefunction fulfills the GP equation
\beqa
\mu \psi_{-3}(\bold{r}) &=&
\bigg[-{\hbar^2\nabla^2\over2M} + V_{\rm trap}(\bold r) -3p +9q\bigg] 
\hat \psi_{-3} (\bold r) \nonumber \\
 &+& 
g  n_{-3}(\bold{r})\psi_{-3}(\bold{r})
\label{eq:sch}  \\ 
&-& 
36\sqrt{\pi\over 5} c_d \int {d\bold{r}'\over |\bold{r}-\bold{r}'|^3}
Y_{20}(\widehat{\bold{r}-\bold{r}'}) \psi_{-3}^2(\bold{r}')\psi_{-3}(\bold{r}) \, ,
\nonumber 
\label{eq:gpd}
\eeqa
with $g\equiv c_0 + 9c_1 + 81c_3$. 

In homogeneous space, $V_{\rm trap}(\bold{r})=0$, $\psi_{-3}(\bold{r})=\psi_{-3}$ and 
$\mu=9q -3p+ g n_{-3}$. 
Moving into momentum space, $\delta\hat\psi_{m}(\bold{k})$, 
we introduce the operators
\beqa
\hat{O}_{-3,\pm}(\bold{k}) &\equiv& \delta\hat{\psi}_{-3}(\bold k) 
\pm \delta\hat{\psi}_{-3}^\dagger(-\bold{k}) \nonumber \\
\hat{O}_{-2,\pm}(\bold{k}) &\equiv &{\rm e}^{-i\phi_k} 
\delta\hat{\psi}_{-2}(\bold k) \pm 
\delta\hat{\psi}_{-2}^\dagger(-\bold{k}) {\rm e}^{i\phi_k} \;,
\eeqa
where we have introduced spherical coordinates $\bold{k}=(k,\theta_k, \phi_k)$.
These operators are governed by a set of coupled Heisenberg equations:
\beqa
i\hbar \left(\begin{array}{c} \dot{\hat{O}}_{-3,+}(\bold{k})\\ 
\dot{\hat{O}}_{-2,+}(\bold{k})\end{array}\right) &=&
\hat{R}(k) \left(\begin{array}{c} \hat{O}_{-3,-}(\bold{k})\\ 
\hat{O}_{-2,-}(\bold{k})\end{array}\right) 
\nonumber \\
i\hbar \left(\begin{array}{c} \dot{\hat{O}}_{-3,-}(\bold{k})\\ 
\dot{\hat{O}}_{-2,-}(\bold{k})\end{array}\right) &=&
\hat{M}(\bold{k})
\left(\begin{array}{c} \hat{O}_{-3,+}(\bold{k})\\ 
\hat{O}_{-2,+}(\bold{k})\end{array}\right) \label{eq:o} \;,
\eeqa
where
\beqa
\hat{R}(k) &=& 
\left(\begin{array}{cc}
{\hbar^2 k^2\over2M} & 0 \\ 0 &
{\hbar^2 k^2\over2M} -U
- 4\pi  c_d n_{-3}
\end{array}\right) \label{rk}\,,
\eeqa
with $U\equiv 5q-p$, 
and 
\beqa
\hat{M}(\bold{k}) &=&
\left(\begin{array}{cc}
M_{11}& M_{12}\\
M_{21}&M_{22}\\
\end{array}\right) \label{mk}
\eeqa
with 
\beqa
M_{11}&=&{\hbar^2 k^2\over2M} + 2 g n_{-3}  
+ 24\pi c_d n_{-3} \big(3\cos^2\theta_k - 1\big)\,, \nonumber\\
M_{12}&=& M_{21}= 36 \pi \sqrt{2\over 3} c_d n_{-3} \sin\theta_k\cos\theta_k \,, \nonumber\\
M_{22}&=& {\hbar^2 k^2\over2M} -U- 4\pi  c_d n_{-3} \big(1 - 3\sin^2\theta_k\big) \,.
\label{matrix}
\eeqa
Note that contrary to the spin-$1$ case, scalar fluctuations (given by $\delta\hat{\Psi}_{-3}$) couple with spin fluctuations (given by $\delta\hat{\Psi}_{-2}$) at first order. 
As a consequence, the elementary excitations have a hybrid scalar/spin character absent in the spin-$1$ case.

The corresponding Bogoliubov excitations may be written as a linear combination of the operators above: 
\beqa
\left(\begin{array}{c} \Lambda_{+}(\bold{k}) \\ \Lambda_{-}(\bold{k}) \\
\Lambda_{+}^\dagger(-\bold{k}) \\ \Lambda_{-}^\dagger(-\bold{k})
\end{array}\right)
&=& 
T(\bold{k}) 
\left(\begin{array}{c} \hat{O}_{-3,-}(\bold{k}) \\ \hat{O}_{-2,-}(\bold{k}) \\
\hat{O}_{-3,+}(\bold{k}) \\ \hat{O}_{-2,+}(\bold{k})
\end{array}\right) \,,
\eeqa
where
\beqa
T(\bold{k}) &=& \left(\begin{array}{cccc}
\alpha_+(\bold{k}) & \beta_+(\bold{k}) & \gamma_+(\bold{k}) & \delta_+ (\bold{k})\\
\alpha_-(\bold{k}) & \beta_-(\bold{k}) & \gamma_-(\bold{k}) & \delta_-(\bold{k}) \\
\alpha_+(\bold{k}) & \beta_+(\bold{k}) & -\gamma_+(\bold{k}) & -\delta_+(\bold{k}) \\
\alpha_-(\bold{k}) & \beta_-(\bold{k}) & -\gamma_-(\bold{k}) & -\delta_-(\bold{k}) \\
\end{array}\right) \,,
\eeqa
and 
\beqa
\alpha_\pm^2(\bold{k}) &=& {B^2(\bold{k})R_{11}(k) R_{22}(k)\over 
4E_\pm(\bold{k})} \Big[B^2(\bold{k})R_{22}(k) 
\nonumber \\ && + \big(E_\pm^2(\bold{k})-A(\bold{k})\big)^2 R_{11}(k) \Big]^{-1}
\nonumber \\
\beta_{\pm}(\bold{k})  &=& \alpha_\pm(\bold{k})  
{E_\pm^2(\bold{k})-A(\bold{k}) \over B(\bold{k})}
\nonumber \\
\gamma_{\pm}(\bold{k}) &=& \alpha_\pm(\bold{k})  {E_\pm(\bold{k}) \over R_{11}(k)}
\nonumber \\
\delta_{\pm}(\bold{k}) &=& \beta_{\pm}(\bold{k})  {E_\pm(\bold{k}) \over R_{22}(k)} \,.
\eeqa
The excitation Hamiltonian acquires then the form
\beqa
\hat{H} &=& \int d\bold{k} \ 
\sum_{\kappa=\pm} E_\kappa (\bold{k}) \hat\Lambda_{\kappa}^\dagger(\bold{k})\hat\Lambda_{\kappa}(\bold{k}) \,.
\eeqa
with eigenenergies: 
\beqa
E_\pm^2(\bold{k})  &=& 
{1\over 2}\big(A(\bold{k})+D(\bold{k})\big) \nonumber \\
&& \pm {1\over 2} \sqrt{\big(A(\bold{k})-D(\bold{k})\big)^2 + 4 C(\bold{k})B(\bold{k})} \label{eq:ev} \,,
\eeqa
where $A(\bold{k})\equiv R_{11}(k) M_{11}(\bold{k})$, $B(\bold{k})\equiv R_{11}(k) M_{21}(\bold{k})$, $C(\bold{k})\equiv R_{22}(k) M_{12}(\bold{k})$, and 
$D(\bold{k})\equiv R_{22}(k) M_{22}(\bold{k})$. The modes are stable, i.e. possess real eigen-energies,  if $U<U_{cr}=-4\pi n_3 c_d/g_6$. Note that stability is just governed by the interplay between Zeeman energies and 
the spin relaxation due to the DDI. 

Defining the matrices:
\beqa
\hat{\Delta}&=& {1\over \alpha_+\beta_--\alpha_-\beta_+}
\left(\begin{array}{cc}
\beta_- & - \beta_+ \\
-\alpha_- & \alpha_+
\end{array} \right) \nonumber \\
\hat{\Gamma}&=& {1\over \gamma_+\delta_--\gamma_-\delta_+}
\left(\begin{array}{cc}
\delta_- & - \delta_+ \\
-\gamma_- & \gamma_+
\end{array} \right) 
\eeqa
we may express the population in $m=-2$ in the form
\beqa
\langle \hat{n}_{-2}(k,\theta_k)\rangle &=& {\tilde A_{2+} \over {\rm e}^{E_+\over k_BT} -1}
+ {\tilde A_{2-} \over {\rm e}^{E_-\over k_BT} -1} + Q_{-2}
\eeqa
where we introduce the amplitudes
\beqa
A_{2+}(\bold{k}) &=& {1\over 8}\bigg[\Gamma_{21}^2(\bold{k})+\Delta_{21}^2(\bold{k})\bigg] ,\\
A_{2-}(\bold{k}) &=& {1\over 8}\bigg[\Gamma_{22}^2(\bold{k})+\Delta_{22}^2(\bold{k})\bigg],
\eeqa
and the zero temperature quantum fluctuations 
\beqa
Q_{-2}(\bold{k}) &=& {1\over 16} 
\Bigg\{\bigg[\Gamma_{21}(\bold{k})+\Delta_{21}(\bold{k})\bigg]^2 \nonumber \\ 
&+& \bigg[\Gamma_{22}(\bold{k})+\Delta_{22}(\bold{k})\bigg]^2 \Bigg\}.
\eeqa

\begin{figure}[t]
\begin{center}
\includegraphics[width=0.95\columnwidth, clip=true]{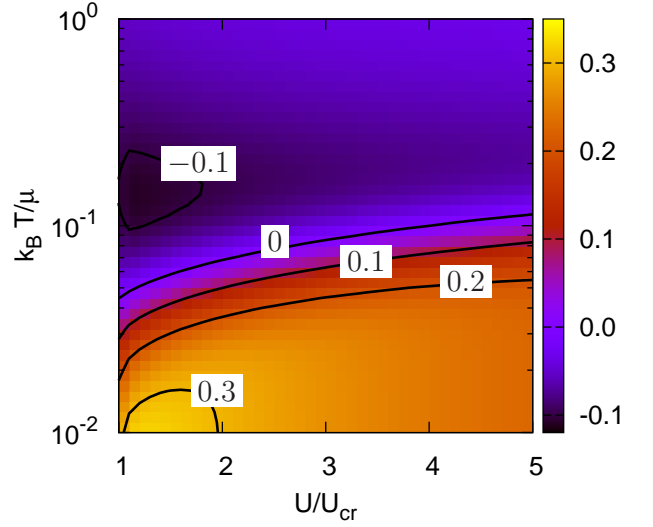}
\caption[]{Anisotropic $\chi$
as a function of $k_BT/\mu$,  and $U/U_{cr}$, for an homogeneous $^{52}$Cr BEC
with $n_{-3}=10^{14}$ particles/cm$^3$. The black lines describe 
configurations with the same anisotropy, the corresponding $\chi$ values are indicated.}  
\label{fig:3}
\end{center}
\end{figure}

Contrary to the spin-$1$ case discussed in the previous section,  $\langle\hat{n}_{-2}\rangle$ has in general an angular dependence, 
which results from the anisotropy of the DDI. We may quantify the anisotropy of the spin population by means of 
$\chi = \int (3\cos^2\theta_k-1) \langle\hat{n}_{-2}(k,\theta_k)\rangle d^3 k$. An isotropic distribution is characterized by $\chi=0$, whereas 
positive values indicate a distribution preferentially oriented along  $\theta_k=\pi/2$. The anisotropy $\chi$ presents an interesting 
dependence as a function of temperature and $U/U_{cr}$, depicted in Fig.~\ref{fig:3} for the case of $n_{-3} = 10^{14}$ cm$^{-3}$, $U_{cr}/\mu_0 = -0.05$.
At low $T$, $\chi$ has small positive values, increasing when $U/U_{cr}$ 
increases~(for very low $T\lesssim 0.01\mu/k_B$, $\chi$ 
acquires a maximum for intermediate $U$ values).
For larger $T$, $\chi<0$ for low $U/U_{cr}$ indicating a momentum distribution oriented along $\theta_k=0$, whereas 
at larger $U/U_{cr}$ the distribution becomes basically isotropic.

We consider at this point the case of a trapped Chromium condensate. 
In general, LDA must be carefully considered, due to the long range character of the DDI. 
However, when the characteristic length of this 
interaction, $a_{dd}=3Mc_d/\hbar$, is much smaller than 
the typical length of the condensate harmonic trap, 
$a_{h.o}=\sqrt{\hbar/M\omega}$, the LDA can still be 
used to calculate the number of particles in the  $m=-2$ state,
as long as the density profile of the  $m=-3$ BEC 
varies smoothly with $r$. This approximation allows to estimate the total number of atoms 
in $m=-2$, but is of course not appropriate to study its angular 
distribution. 

For an axially symmetric harmonic potential 
$V_{\rm trap}(\bold r) = M\omega_\perp^2(r_\perp^2 + \lambda^2 z^2)/2$, 
where $\lambda = \omega_z/\omega_\perp$ is the trap anisotropy. 
The density profile is,
\beqa
n(r_\perp,z) = {15 N\over 8\pi R_\perp^2 R_z} 
\bigg(1-{r_\perp^2\over R_\perp^2}-{z^2\over R_z^2}\bigg)
\eeqa
where $\kappa=R_\perp/R_z$ is found by solving~\cite{odell,eberlein}
\beqa
{\kappa^2\over \lambda^2}\bigg[3\varepsilon_{dd} 
{f(\kappa)\over 1-\kappa^2}
\bigg({\lambda\over 2}+1\bigg) - 2\varepsilon_{dd} - 1\bigg] = \varepsilon_{dd} - 1 \,  ,
\eeqa
with $\varepsilon_{dd}= 12\pi^2 c_d /g$, and
\beqa
f(\kappa) = {1+2\kappa^2\over 1-\kappa^2} 
- {3\kappa^2\over (1-\kappa^2)^{3/2}} \tanh^{-1}\sqrt{1-\kappa^2} \, .
\eeqa

Normalizing the total density to the total number of particles, one finds:
\beqa
R_\perp = \bigg[\alpha \kappa \bigg\{1-\varepsilon_{dd}
\bigg(1-{3\over 2}{\kappa^2\over 1-\kappa^2}f(\kappa)\bigg)\bigg\}\bigg]^{1/5}
\eeqa
where $\alpha\equiv 15gN/4\pi M \omega_\perp^2$. 
The  chemical potential is $\mu = gn(0,0) \Big(1-\varepsilon_{dd} f(\kappa)\Big)$.
At each small volume, the local chemical potential is 
$\mu(\bold{r})=\mu - V_{\rm trap}(\bold{r})$, which may be well approximated for Chromium by  
$\mu(r,z) \simeq g n_{-3}(r,z)$, as for the homogeneous BEC. 

\begin{figure}[t]
\includegraphics[width=0.95\columnwidth, clip=true]{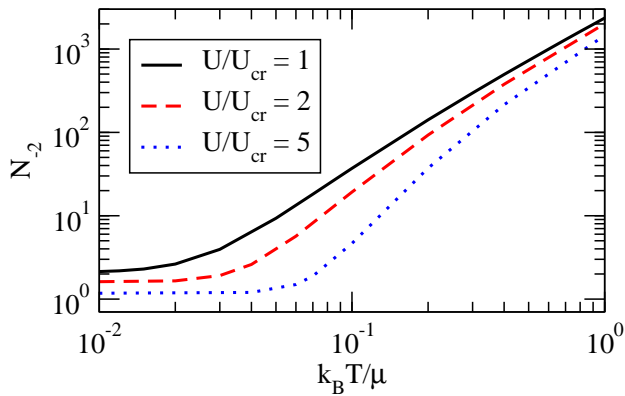}
\caption[]{Number of Chromium atoms in $m=-2$ as a function 
of $k_BT/\mu$ for different values of $U/U_{cr}$, for the case discussed in the text.}  
\label{fig:4}
\end{figure}

Fig.~\ref{fig:4} shows $N_{-2}$ as a function of temperature for several values of the $U/U_{cr}$, for 
the specific case of a $^{52}$Cr BEC with $N=10^5$ in $m=-3$ in a 
spherical trap of frequency $\omega = 2\pi \times 50$Hz, where $U_{cr}$ is determined by the central 
density $n(0,0)=5.33\cdot 10^{13}$cm$^{-3}$. Note that $a_{dd} = 0.894$nm $\ll a_{h.o.} = 1.97\mu$m, and hence 
well within the limits of the LDA. Close to $U_{cr}$ populations of $N_{-2}=10$ may be attained below $0.1 \mu/k_B$. Hence, 
as for the spin-$1$ case, Fig.~\ref{fig:4} shows clearly that one may employ the population 
in $m=-2$ (combined with an abrupt jump into instability, as discussed for spin-$1$)  for thermometry purposes.
Finally, we should note that as for the case of spin-$1$ isotropic curves bend towards lower $T$ when approaching $U_{cr}$, and hence 
also for Chromium an adiabatic reduction of $U/U_{cr}$ may allow for an interesting cooling mechanism. 

\section{Conclusions}
\label{sec:Conclusions}

We have analyzed the thermal activation of spin excitations in spinor condensates. 
For the case of spin-$1$ condensates, an stable $m=0$ condensate presents a non-negligible 
thermally activated population of $m=\pm 1$ due to spin-changing collisions. For the case of a stable Chromium BEC 
in $m=-3$ dipole-induced spin-relaxation leads as well to thermal activation, which contrary to the spin-$1$ case, acquires an intriguing 
temperature-dependent anisotropy. For both cases we have shown that the spin population may be employed at very low $T\ll \mu/k_B$ 
as a possible mechanism for deep-temperature thermometry. We have shown as well that an external adiabatic variation of the Zeeman energy 
may be employed to achieve an adiabatic cooling mechanism.

\begin{acknowledgments}
We thank Bruno Laburthe-Tolra and Carsten Klempt for interesting
discussions. We acknowledge support from the Spanish MICINN grants FIS2011-24154 and FIS2008-00784 (TOQATA), Generalitat de Catalunya (2009-SGR1289), and the Cluster of Excellence QUEST. M. M.-M. is supported by an FPI PhD grand of the Ministerio de Ciencia e Innovaci\'on (Spain). B. J.-D. is supported by the Ram\'on y Cajal program.
\end{acknowledgments}

\end{document}